\documentclass[10pt,twocolumn,showpacs,pra,superscriptaddress]{revtex4-1}

\usepackage[tbtags]{amsmath}
\usepackage{amsthm}
\usepackage{amssymb}
\usepackage{amsfonts}
\usepackage{cases}
\usepackage{bbm,bm}
\usepackage{bbding}
\usepackage{graphicx}
\usepackage{xcolor}
\usepackage{hyperref}
\hypersetup{colorlinks=true,breaklinks=true,pdfstartview=FitH,linkcolor=blue,citecolor=brown}

\allowdisplaybreaks

\DeclareMathOperator{\Tr}{Tr}%
\DeclareMathOperator{\diag}{diag}%

\newcommand{\bra}[1]{\langle #1 \vert}
\newcommand{\ket}[1]{\vert #1 \rangle}

\theoremstyle{plain}
\newtheorem{lemma}{Lemma}

\begin{document}
\title{Symmetric geometric measure and dynamics of quantum discord}

\author{Mingjun Shi}%
\email{shmj@ustc.edu.cn}%
\affiliation{Department of Modern Physics, University of Science and Technology of China, Hefei, Anhui 230026, People's Republic of China}

\author{Fengjian Jiang}
\email{jfjiang@mail.ustc.edu.cn}
\affiliation{Department of Modern Physics, University of Science and Technology of China, Hefei, Anhui 230026, People's Republic of China}
\affiliation{Huangshan University, Huangshan, Anhui 245021, People's Republic of China}

\author{Jiangfeng Du}
\affiliation{Department of Modern Physics, University of Science and Technology of China, Hefei, Anhui 230026, People's Republic of China}

\begin{abstract}
A symmetric measure of quantum correlation based on the Hilbert-Schmidt distance is presented in this paper. For two-qubit states, we simplify considerably the optimization procedure so that numerical evaluation can be performed efficiently. Analytical expressions for the quantum correlation are attained for some special states. We further investigate the dynamics of quantum correlation of the system qubits in the presence of independent dissipative environments. Several nontrivial aspects are demonstrated. We find that the quantum correlation can increase even if the system state is suffering dissipative noise. Sudden changes occur, even twice, in the time evolution of quantum correlation. There is certain correspondence between the evolution of quantum correlation in the systems and that in the environments, and the quantum correlation in the systems will be transferred into the environments completely and asymptotically.
\end{abstract}


\maketitle

\section{introduction}

Quantum systems can be correlated in ways inaccessible to classical objects.
There are quantum states that cannot be prepared with the help of local operations and classical communication, or that cannot be represented as a mixture of product states.
These states are called entangled states.
Entanglement is certainly a kind of quantum correlation, and moreover is by far the most famous and best studied kind of quantum correlation \cite{Horodecki.RevModPhys.81.865.2009}.
One reason for this situation is the fact that quantum entanglement plays an important role in much of the research of quantum information science \cite{Nielsen.QCandQI.2000}.
However, quantum entanglement is not the only aspect of nonclassicality of correlations and does not account for all nonclassical properties of quantum correlations.
A typical example is that quantum nonlocality can arise without entanglement \cite{Bennett.PhysRevA.59.1070.1999}.
More importantly, although it is well known that entanglement is essential for certain kinds of quantum-information
tasks like teleportation \cite{Bennett.PhysRevLett.70.1895.1993} and super-dense coding \cite{Bennett.PhysRevLett.69.2881.1992}, there is no definite answer as to whether all quantum algorithms that outperform their best known classical counterparts require entanglement as a resource.
Indeed, there are several instances where we see a quantum improvement in the absence or near absence of entanglement (see e.g. \cite{Braunstein.PhysRevLett.83.1054.1999,Meyer.PhysRevLett.85.2014.2000,
Vidal.PhysRevLett.91.147902.2003,Biham.TheorCompSci.320.15.2004}).
In particular, no entanglement is present in the computational model referred to as ``the power of one qubit'' with the acronym DQC1 \cite{Knill.PhysRevLett.81.5672.1998,Laflamme.QuanInfoComput.2.166.2002}. Despite this, the mixed separable states can create an advantage for computational tasks over their classical counterparts
\cite{Laflamme.QuanInfoComput.2.166.2002,
      Datta.PhysRevA.72.042316.2005,
      *Datta.PhysRevA.75.042310.2007}.

On the other hand, many works have been devoted to understanding and quantifying the quantum correlations beyond entanglement
\cite{Henderson.JPhysA.34.6899.2001,Ollivier.PhysRevLett.88.017901.2001,
      Oppenheim.PhysRevLett.89.180402.2002,Koashi.PhysRevA.69.022309.2004,
      Groisman.PhysRevA.72.032317.2005,Horodecki.PhysRevA.71.062307.2005,
      Luo.PhysRevA.77.022301.2008,
      Rodriguez.41.205301.2008,*Shabani.PhysRevLett.102.100402.2009,
      Piani.PhysRevLett.100.090502.2008,*Piani.PhysRevLett.102.250503.2009,
      Modi.PhysRevLett.104.080501.2010,Dakic.PhysRevLett.105.190502.2010,
      Adesso.PhysRevLett.105.030501.2010,Luo.PhysRevLett.106.120401.2011,
      Cavalcanti.PhysRevA.83.032324.2011,*Madhok.PhysRevA.83.032323.2011,
      Streltsov.PhysRevLett.106.160401.2011,*Piani.PhysRevLett.106.220403.2011}.
As the total correlation can be split into a classical part and a quantum part  \cite{Henderson.JPhysA.34.6899.2001},
various measures of quantum correlations are proposed by considering different notions of classicality and operational means to quantify quantumness.
Amongst them, quantum discord \cite{Ollivier.PhysRevLett.88.017901.2001} has attracted much attention. Quantum discord has been defined as the mismatch between two quantum analogues of classically equivalent expression of the mutual information. It can be also expressed as the difference between total correlation, measured by quantum mutual information, and the classical correlation defined in \cite{Henderson.JPhysA.34.6899.2001}. The notion of quantum discord goes beyond entanglement: separable states can have nonzero discord. In particular, it is believed that quantum discord is the figure of merit for the DQC1 model of quantum computation \cite{Datta.PhysRevLett.100.050502.2008}.
Generally speaking, quantum discord plays an important role in quantum information processing \cite{Merali.Nature.474.24.2011}.

In order to obtain the classical correlation (and thereby the quantum discord) of a bipartite quantum system, one has to perform measurement on one subsystem to extract the information about the other subsystem, i.e., locally accessible information. Hence this sort of one-sided measurements implies that classical correlation and quantum discord are not symmetric under the permutation of subsystems. As a measure of correlation, whether classical or nonclassical, one would expect that it is symmetric.
Some symmetric measures of correlations have been proposed in Refs. \cite{Terhal.JMathPhys.43.4286.2002,Horodecki.PhysRevA.71.062307.2005,
      Luo.PhysRevA.77.022301.2008,
      Modi.PhysRevLett.104.080501.2010}.
The key step in obtaining these measures is to consider the classical-classical (CC) states.
The state $\rho$ is called a CC state, if it can be expressed as the mixture of locally distinguishable states, namely, $\rho_{cc}=\sum_{i,j}p_{ij}\ket{i}\bra{i}\otimes\ket{j}\bra{j}$,
where $p_{ij}$ is a joint probability distribution and local states $\ket{i}$ and $\ket{j}$ span an orthonormal basis. The set of product states is the subset of the set of CC states, while the set of CC states in turn is the subset of the set of separable states. A CC state is in fact the embedding of a classical probability distribution in the formalism of quantum theory and as such has no quantumness.
Given a state $\rho$, one can find the closest CC state, $\rho_{cc}$, to it. It is natural to regard this minimal distance as the measure of quantum correlation. Such a measure has a transparent geometrical meaning. The distance can be measured with trace distance, Hilbert-Schmidt distance, Bures distance, relative entropy and so on.
In \cite{Modi.PhysRevLett.104.080501.2010}, the authors used the relative entropy as the distance measure to provide the unified view of quantum and classical correlation.

We present in this paper a symmetric geometric measure of quantum correlation based on Hilbert-Schmidt (HS) distance.
By applying von Neumann measurement on each subsystem, any bipartite state will become a CC state, which we call a measurement-induced CC (MICC) state.
Given a bipartite state $\rho$, we define the quantum correlation, denoted by $G(\rho)$, as the squared HS distance between $\rho$ and the closest MICC state.
The evaluation of $G(\rho)$ requires an optimization procedure over the set of all local von Neumann measurements on two subsystems, and thus attacking the general case is a formidable task.
However, for 2-qubit states, we are able to simplify the optimization over two-sided measurements to that over one-sided measurements.
In other words, the number of measurement parameters over which the optimization procedure is performed is reduced from four to two.
Thus we are able to evaluate the geometric measure of quantum correlation by the efficient numerical method.
We further find the exact analytical expressions for some special states.

Moveover, we will discuss the dynamics of quantum correlation, quantified by the geometric measure $G$, when quantum states undergo a noisy channel.
It has been shown that the evolution of quantum correlation, whether in terms of quantum entanglement or of quantum discord, may behave in a ``sudden'' way.
Quantum entanglement can evolve to sudden death or birth 
\cite{Yu.PhysRevLett.93.140404.2004,
      *Eberly.Sci.316.555.2007,
      *Yu.Sci.323.598.2009}.
The decoherence regime of quantum discord may change suddenly to that of classical correlation \cite{Mazzola.PhysRevLett.104.200401.2010}.
Also, the dynamics of quantum discord has attracted much attention 
\cite{Maziero.PhysRevA.80.044102.2009,*Maziero.PhysRevA.81.022116.2010,
      Werlang.PhysRevA.80.024103.2009,Wang.PhysRevA.81.014101.2010,
      *Fanchini.PhysRevA.81.052107.2010}.
This ``sudden'' behavior has not been thoroughly understood yet.
It is natural to enquire as to whether and how such a geometric measure of quantum correlation will exhibit some sort of sudden changes.
To this end, we consider the case that a 2-qubit state is effected by the action of two independent non-unital channels, e.g., amplitude damping (AD) channel, and calculate analytically $G(t)$, the time evolution of the geometric measure developed in this paper.
We observe the nontrivial phenomenon that the function $G(t)$ may be neither monotone nor smooth.
In other words, during the time evolution, the quantum correlation can increase under the influence of AD channel, and the rate of evolution can exhibit sudden changes at some critical times.
Despite these novel aspects, the asymptotical behavior of $G(t)$ is in accordance with the coherence decay.
In addition, we see that the quantum correlation in the system qubits is completely transferred to the environments after sufficiently long time.


\section{Definitions and notations}

This section is a prelude providing the definitions and notations that will be used throughout the whole paper.

\subsection{Hilbert-Schmidt distance}

For any 2-qubit state $\rho$ shared by two parties $A$ and $B$, let's define a $4\times4$ matrix $R$ as $R=(R_{\mu\nu})$ with the elements $R_{\mu\nu}$ given by $R_{\mu\nu}=\Tr[\rho(\sigma_\mu\otimes\sigma_\nu)]$ for $\mu,\nu=0,1,2,3$, where $\sigma_0$ is $2\times2$ identity matrix and $\sigma_i$ ($i=1,2,3$) are usual Pauli matrices.
We write $R$ as
\begin{equation} \label{R matrix}
  R=
  \begin{pmatrix}
    1 & \vec{y} \\
    \vec{x}^{\,T} & T
  \end{pmatrix},
\end{equation}
where $\vec{x}$ and $\vec{y}$ are the Bloch vectors (in row form) of reduced state $\rho^A$ and $\rho^B$ respectively, and $T$ is a $3\times3$ matrix and usually called correlation matrix.
The superscript $T$ denotes transposition.

Hilbert-Schmidt (H-S) norm (also called Frobenius norm) is defined as
$\Vert M\Vert_{\mathrm{HS}}=\sqrt{\Tr(MM^\dag)}$
for any bounded operator $M$ on complex-dimensional Hilbert space.
It follows that the H-S distance between two operators $M$ and $N$ is given by
\begin{equation}
  D_{\mathrm{HS}}(M,N)=\sqrt{\Tr[(M-N)(M^\dag-N^\dag)]}.
\end{equation}
In the following, we will omit the subscript $\mathrm{HS}$.

Let $R_\rho$ and $R_\tau$ be the $R$ matrix associated with 2-qubit states $\rho$ and $\tau$ respectively. The H-S distance between $\rho$ and $\tau$ can be expressed in terms of the elements of $R$ matrix:
\begin{equation}\label{HS distance}
  D^2(\rho,\tau)=\frac{1}{4}\big[|\vec{x}_\rho-\vec{x}_\tau|^2
    +|\vec{y}_\rho-\vec{y}_\tau|^2+\Vert T_\rho-T_\tau\Vert^2\big].
\end{equation}

Now suppose that we perform local von Neumann measurements on both qubits $A$ and $B$. The measurement operators for $A$ and $B$ are given by
\begin{equation*}
    \Pi^A_{\pm}=\frac{1}{2}(\mathbbm1\pm \vec{k}\cdot\vec{\sigma}), \quad
    \Pi^B_{\pm}=\frac{1}{2}(\mathbbm1\pm \vec{\ell}\cdot\vec{\sigma}),
\end{equation*}
respectively,
where $\vec{k}$ and $\vec{\ell}$ are unit vectors in three-dimensional real space.
After measurements, we obtain a MICC state $\chi$, that is,
\begin{equation}
  \chi=\sum_{i,j=+,-}(\Pi^A_i\otimes\Pi^B_j)\,\rho\,(\Pi^A_i\otimes\Pi^B_j)
\end{equation}
Let $K=\vec{k}^{\,T}\,\vec{k}$ and $L=\vec{\ell}^{\,T}\,\vec{\ell}$. Both of them are $3\times3$ real symmetric matrices.
The $R$ matrix of $\chi$ can be written as
\begin{equation}
  R_{\chi}=
  \begin{pmatrix}
    1 & \vec{y}\,L \\ K\,\vec{x}^{\,T} & KTL
  \end{pmatrix}.
\end{equation}
From \eqref{HS distance}, the squared distance between $\rho$ and $\chi$ is given by
\begin{equation*}
\begin{split}
  & D^2(\rho,\chi) \\
  = & \frac{1}{4}\big[|\vec{x}-\vec{x}K|^2
      +|\vec{y}-\vec{y}L|^2+\Vert T-KTL\Vert^2\big]  \\
  = & \frac{1}{4}\Big\{x^2+y^2+\Vert T\Vert^2
      -\big[\vec{x}K\vec{x}^{\,T}+\vec{y}L\vec{y}^{\,T}+\Tr(TLT^TK)\big]\Big\},
\end{split}
\end{equation*}
where $x^2=\vec{x}\cdot\vec{x}$ and so forth, and in the last line we have used $K^2=K$ and $L^2=L$. Let $X=\vec{x}^{\,T}\,\vec{x}$ and $Y=\vec{y}^{\,T}\,\vec{y}$.
We rewrite $D^2(\rho,\chi)$ as
\begin{equation}\label{Dsq between rho and rhocc}
\begin{split}
  D^2(\rho,\chi)= & \frac{1}{4}\Big[\Tr(X+Y+TT^T) \\
    &  \qquad -\Tr\big(XK+YL+TLT^TK\big)\Big].
\end{split}
\end{equation}

If only one qubit, say $A$, is measured, the resulting state is a classical-quantum (CQ) state.
A general CQ state has the form of $\sum_ip_i\ket{i}\bra{i}\otimes\rho_i$.
In the case we are considering, the measurement-induced CQ state is expressed as
\begin{equation} \label{CQ state}
  \rho^{\to}=(\Pi_+^A\otimes\mathbbm1)\,\rho\,(\Pi_+^A\otimes\mathbbm1)
    +(\Pi_-^A\otimes\mathbbm1)\,\rho\,(\Pi_-^A\otimes\mathbbm1).
\end{equation}
The corresponding $R$ matrix is
\begin{equation}
  R_{\rho^\to}=
  \begin{pmatrix}
    1 & \vec{y} \\ K\,\vec{x}^{\,T} & KT
  \end{pmatrix}.
\end{equation}
It follows that the squared distance between $\rho$ and $\rho^{\to}$ is given by
\begin{equation}\label{Dsq between rho and rhocq}
  D^2(\rho,\rho^{\to})
  = \frac{1}{4}\big[\Tr(X+T^TT)-\Tr(XK+TT^TK)\big].
\end{equation}
If only qubit $B$ is measured, the squared distance $D^2(\rho,\rho^{\leftarrow})$ can be defined similarly.

\subsection{Quantum channel}

We will give only a brief introduction to the quantum channel, and refer the reader to books, say, \cite{Nielsen.QCandQI.2000,Bengtsson.GeoQuantState.2006} for a detailed analysis.

A quantum channel is a trace preserving completely positive (TPCP) map. A linear map $\mathcal{E}$ is completely positive if and only if it is of the form
\begin{equation} \label{Kraus sum}
    \rho\,\longrightarrow\,\rho'=\sum_iK_i\rho K^\dag_i,
\end{equation}
where $K_i$ are the operators on the state space of the system and are known as Kraus operator.
Eq. \eqref{Kraus sum} is called operator-sum representation of a quantum operation.
A completely positive map is called trace-preserving if and only if
\begin{equation} \label{TPCP cond}
  \sum_iK_i^\dag K_i=\mathbbm1.
\end{equation}
If a quantum channel leaves the maximal mixed state invariant, it is called a bistochastic map or unital channel, that is, $\sum_iK_iK_i^\dag=\mathbbm1$.
Otherwise, the channel is non-unital.

The action of a quantum channel on a quantum system can be described by a unitary evolution on an extended system, the system $S$ plus the environment $E$, followed by a partial trace over $E$.
This description is called the unitary representation of the quantum channel.
This representation is not unique since many different unitary evolutions will lead to the same effect of the channel.

Now let us consider a concrete quantum channel, amplitude damping (AD) channel.
The AD channel is used to describe the evolution of the system state in the presence of a dissipative environment.
In this process the system interacts with a thermal bath at zero temperature.
This process could be described as spontaneous emission of a two-state atom (system $S$) coupled with the vacuum modes of the ambient electromagnetic field (environment $E$) which leads the atom state to the ground state (see, e.g., \cite{Breuer.OpenQuantSys} for more in-depth discussion).

Considering the behavior of a two-level atom in a $N$-mode cavity, the interaction between them gives rise to the phenomenological map
\begin{align}
  & \ket{0^S}\ket{\mathbf{0}^E}
    \longrightarrow\ket{0^S}\ket{\mathbf{0}^E}, \label{AD map1}\\
  & \ket{1^S}\ket{\mathbf{0}^E}\longrightarrow
    \gamma(t)\ket{1^S}\ket{\mathbf{0}^E}
      +\sqrt{1-\gamma^2(t)}\,\ket{0^S}\ket{\mathbf{1}^E}.
    \label{AD map2}
\end{align}
Here, $\ket{0^S}$ is the ground state and $\ket{1^S}$ is the excited state of the atom,
while $\ket{\mathbf{0}^E}$ is the vacuum state of the cavity and $\ket{\mathbf{1}^E}$ describe the cavity state with only one excitation distributed over all modes. The amplitude $\gamma(t)$ converges to $\gamma(t)=\exp(-\kappa t/2)$ in the limit of $N\to\infty$. 
The parameter $\kappa$ is usually called coupling strength. 
Then the atom and environment evolve as an effective 2-qubit system.

From \eqref{AD map1} and \eqref{AD map2}, we can write the Kraus operators of the AD channel:
\begin{equation}
  K_0=
    \begin{pmatrix}
      1 & 0 \\ 0 & \gamma
    \end{pmatrix}, \quad
  K_1=
    \begin{pmatrix}
      0 & \sqrt{1-\gamma^2} \\ 0 & 0
    \end{pmatrix}.
\end{equation}
The evolution of the system state is then given by
$\rho^S(t)=\sum_{i=0}^{1}K_i\rho^SK_i^\dag$.

Let us further consider the case of a 2-qubit system being affected by their independent dissipative environments. We assume that at time $t=0$ the system-plus-environment state is described by
\begin{equation}
  \rho^{ABA'B'}=\rho^{AB}\otimes
    \ket{\mathbf{0}^{A'}\mathbf{0}^{B'}}\bra{\mathbf{0}^{A'}\mathbf{0}^{B'}},
\end{equation}
where $\ket{\mathbf{0}^{A'}\mathbf{0}^{B'}}$ is the vacuum state of two environments.

By means of Eqs. \eqref{AD map1} and \eqref{AD map2}, we can work out $\rho^{ABA'B'}(t)$,
the total state of $ABA'B'$ at time $t$.
Tracing out environments $A'$ and $B'$, we obtain the system state $\rho^{AB}(t)$, which can be expressed in the operator-sum representation,
\begin{equation} \label{sys state}
  \rho^{AB}(t)=\sum_{i,j=0}^{1}(K_i\otimes K_j)\rho^{AB}
    (K_i^\dag\otimes K_j^\dag).
\end{equation}
On the other hand, the reduced environment state is given by
\begin{equation} \label{env state}
  \rho^{A'B'}(t)=\Tr_{AB}\rho^{ABA'B'}(t).
\end{equation}

\section{Geometric measures of quantum correlation}

As stated in the Introduction, we can define the quantum correlation in a 2-qubit state $\rho$ as the geometric distance, measured by HS norm, between $\rho$ and the closest MICC state $\chi$.
We will present below a detailed analysis in this direction.

\subsection{One-sided measure}

To begin with, we discuss a simpler case in which only one-sided measurement is performed on qubit $A$.
The geometric measure of quantum correlation based on such a sort of one-sided measurements has been proposed in \cite{Dakic.PhysRevLett.105.190502.2010} and discussed in \cite{Luo.PhysRevA.82.034302.2010}.
We recover these results below for completeness.

Given a 2-qubit state $\rho$, one-sided von Neumann measurement on $A$ induces the CQ state $\rho^\to$ given by \eqref{CQ state}.
With the squared distance $D^2(\rho,\rho^\to)$ given by \eqref{Dsq between rho and rhocq}, the geometric measure of quantum correlation is defined as
\begin{equation}
  G^\to(\rho)=\min D^2(\rho,\rho^\to),
\end{equation}
where the minimization is performed over all von Neumann measurements on qubit $A$.

For convenience, we use Dirac bra-ket notation to express the vectors in real three-dimensional space, that is,
$\vec{v}^{\,T}\equiv\ket{v}$ and $\vec{v}\equiv\bra{v}$.
We will use alternatively both notations in the present work.

It can be seen from \eqref{Dsq between rho and rhocq} that to find the minimal value of $D^2(\rho,\rho^\to)$ is equivalent to calculate the maximal value of $\Tr(XK+TT^TK)$.
By noting that $\Tr(XK+TT^TK)=\bra{k}(X+TT^T)\ket{k}$, we have that among all unit $\ket{k}\in\mathbb{R}^3$,
\begin{equation}
    \bra{k}X+TT^T\ket{k}_{\max}=\lambda_{\max},
\end{equation}
where $\lambda_{\max}$ is the largest eigenvalue of $X+TT^T$.
Thus we have
\begin{equation} \label{Dakic measure}
  G^{\to}(\rho)=\frac{1}{4}[\Tr(X+TT^T)-\lambda_{\max}].
\end{equation}
It is exactly the result given in \cite{Dakic.PhysRevLett.105.190502.2010}. Similarly, if qubit $B$ is measured, we have
\begin{equation}
  G^{\leftarrow}(\rho)=\frac{1}{4}[\Tr(Y+T^TT)-\kappa_{\max}],
\end{equation}
with $\kappa_{\max}$ the largest eigenvalue of $Y+T^TT$.
Generally, $G^{\to}\neq G^{\leftarrow}$.

In \cite{Dakic.PhysRevLett.105.190502.2010}, the authors have used general, rather than measurement-induced, CQ state in the derivation of \eqref{Dakic measure}.
Subsequently, it is pointed out in \cite{Luo.PhysRevA.82.034302.2010} that such a geometric measure is essentially measurement-oriented, meaning that the optimal CQ state in the most general sense is indeed the sort of measurement-induced CQ state.

\subsection{Two-sided measure}

Now we proceed to the case of two-sided measurements, which will lead us to a symmetric measure of quantum correlation.
Two-sided measurements will result in MICC states.
The squared HS distance $D^2(\rho,\chi)$ between the state $\rho$ and the corresponding MICC state $\chi$ has been given by \eqref{Dsq between rho and rhocc}.
We define the geometric measure of quantum correlation, based on two-sided measurement, as
\begin{equation}
  G(\rho)=\min_{\{\ket{l},\ket{\ell}\}}D^2(\rho,\chi),
\end{equation}
where the minimization is performed over all $\chi$, or equivalently, over all measurement directions $\{\vec{k},\vec{\ell}\,\}$.
As a CC state does not incorporate any quantum correlation, the two-sided quantifier $G$ can be regarded as a more strict geometric measure of quantum correlation.

By referring to \eqref{Dsq between rho and rhocc}, we need only focus on the maximal value of $\Tr(XK+YL+TLT^TK)$ for all allowed $K$ and $L$.
Let's define the following two matrices:
\begin{align}
  & M=X+TLT^T+\bra{\ell}Y\ket{\ell}\mathbbm1_3, \\
  & N=Y+T^TKT+\bra{k}X\ket{k}\mathbbm1_3,
\end{align}
where $\mathbbm1_3$ denotes the $3\times3$ identity matrix.
It follows that
\begin{equation}
  \Tr(XK+YL+TLT^TK)=\bra{k}M\ket{k}=\bra{\ell}N\ket{\ell}.
\end{equation}
Then the problem is to calculate the maximal value of $\bra{k}M\ket{k}$ or $\bra{\ell}N\ket{\ell}$ over all unit vectors $\ket{k},\ket{\ell}\in\mathbb{R}^3$.

By noting that the eigenvalue of $M$ is equal to the sum of $\bra{\ell}Y\ket{\ell}$ and the eigenvalue of $X+TLT^T$, we see that the eigenvalue of $M$ is the function of $\vec{\ell}$, which we denote by $\lambda_M(\vec{\ell}\,)=\lambda_M(\ell_1,\ell_2,\ell_3)$.
Let $\lambda_M^{\max}$ be the maximal one over all unit vector $\vec{\ell}$.
It follows that the maximal value of $\bra{k}M\ket{k}$ is exactly the $\lambda_M^{\max}$.
Similarly, let $\lambda_N(\vec{k}\,)$ be the eigenvalue of $N$ and $\lambda_N^{\max}$ the maximal one.
The maximal value of $\bra{\ell}N\ket{\ell}$ is $\lambda_N^{\max}$.
Also, we have $\lambda_M^{\max}=\lambda_N^{\max}$.
The directions of the optimal measurement, denoted by $\vec{k}^{\;\mathrm{opt}}$ and $\vec{\ell}^{\;\mathrm{opt}}$, are the eigenvectors of $M$ and $N$ referring to the eigenvalues $\lambda_N^{\max}$ and $\lambda_N^{\max}$, respectively.
Thus we see that the problem of two-sided optimization is indeed reduced to that of the one-sided optimization.

To proceed further, we will use the following lemma, which can be proved by direct calculation.

\begin{lemma} \label{lamma 1}
For any two vectors $\ket{a}$ and $\ket{b}$ (not necessarily normalized) in $\mathbb{R}^3$, the largest eigenvalue of the matrix $\ket{a}\bra{a}+\ket{b}\bra{b}$ is given by
\begin{equation}
    \lambda=\frac{1}{2}\big[a^2+b^2+\sqrt{(a^2-b^2)^2+4\langle a|b\rangle^2}\,\big],
\end{equation}
with $a^2=\langle a|a\rangle$ and $b^2=\langle b|b\rangle$.
The corresponding normalized eigenvector is
\begin{equation*}
\begin{split}
    \ket{\lambda}=\frac{1}{N_\lambda}\Big[\big(a^2-b^2
      +\sqrt{(a^2-b^2)^2+4\langle a|b\rangle^2}\,\big)\,\ket{e_a} & \\
        +\frac{2b}{a}\langle a|b\rangle\,\ket{e_b} & \Big],
\end{split}
\end{equation*}
where $\ket{e_a}$ and $\ket{e_b}$ are the unit vector along $\ket{a}$ and $\ket{b}$ respectively, and $N_{\lambda}$ is the normalization factor.
\end{lemma}

Let $\ket{k'}=T^T\ket{k}$ and $\ket{\ell'}=T\ket{\ell}$. It follows from the Lemma \ref{lamma 1} that
\begin{align}
\begin{split}
  \lambda_M(\vec{\ell}\,)=& \frac{1}{2}\Big[2\bra{\ell}Y\ket{\ell}+x^2+\ell^{\prime\,2} \\
      & \qquad +\sqrt{(x^2-\ell^{\prime\,2})^2+4\langle x|\ell'\rangle^2}\,\Big] \label{lambdaM}
\end{split} \\
\begin{split}
  \lambda_N(\vec{k}\,)=& \frac{1}{2}\Big[2\bra{k}X\ket{k}+y^2+k^{\prime\,2} \\
     & \qquad +\sqrt{(y^2-k^{\prime\,2})^2+4\langle y|k'\rangle^2}\,\Big].
      \label{lambdaN}
\end{split}
\end{align}
The remaining problem is to find the maximal values $\lambda_M^{\max}$ or $\lambda_N^{\max}$.
The geometric measure is then given by
\begin{equation}\label{geo measure}
  G(\rho)=\frac{1}{4}\Big[x^2+y^2+\Vert T\Vert^2-\lambda_{M(N)}^{\max}\Big].
\end{equation}
Referring to \eqref{lambdaM}, we see that $\lambda_M(\vec{\ell}\,)$ involves two parameters, that is, two angles indicating the direction of $\vec{\ell}$.
It is not difficult to attain the $\lambda^{\max}_M$ by efficient numerical method.
For some special states, the exact results are available, which will be stated as follows.

\subsubsection{States with two-sided maximally mixed marginals}

In this case, $\vec{x}=\vec{y}=0$.
We can always transform the correlation matrix $T$ into the diagonal form $\diag(t_1,t_2,t_3)$ by local unitary operations.
Eq. \eqref{lambdaM} becomes $\lambda_M(\vec{\ell}\,)=\ell^{\prime\,2}=\sum_{i=1}^3t_i^2\ell_i^2$. The maximum is given by $\lambda_{M}^{\max}=\max\{t_1^2,t_2^2,t_3^2\}$.

In this case there is no difference between the one-sided measure $G^{\leftrightharpoons}(\rho)$ and the two-sided measure $G(\rho)$.

\subsubsection{States with one-sided maximally mixed marginal}

In this case, only one reduced state, say $\rho^A$, is maximally mixed. Then $\vec{x}=0$. It follows from \eqref{lambdaM} that $\lambda_M(\vec{\ell}\,)=\bra{\ell}TT^T+Y\ket{\ell}$ and the $\lambda_M^{\max}$ is given by the largest eigenvalue of the matrix $TT^T+Y$. On the other hand, if $\rho^{B}$ is maximally mixed, we can consider $\lambda_N(\vec{k})$ with $\vec{y}=0$. Similar results are easily obtained.

\subsubsection{X states with the identical local purity} \label{Sec:X states x3 eq y3}

We call a 2-qubit state the X state if the only nonzero elements in the density matrix lie along the diagonal or skew diagonal.
By local unitary transformations, the entries in the $R$ matrix can always be written as
\begin{align*}
    & \vec{x}=(0,0,x_3), \quad \vec{y}=(0,0,y_3), \\
    & T=\diag(t_1,t_2,t_3).
\end{align*}
The exact expression of $G(\rho)$ for any X state is in fact available.
However, the derivation is too lengthy and too tedious.
So we present the results referring to a restricted class of X states, that is,
the X states with $|x_3|=|y_3|$, i.e., with the identical local purity.
The concrete analytical results are presented in Appendix.


\section{Dynamics of quantum correlation}

Now we are in the position to analyze the dynamics of quantum correlation, in terms of the geometric measure developed in this paper.
We focus on a non-unital channel, AD channel.
It is assumed that the identical AD channel is applied on each qubit independently and simultaneously.
In order for the analysis to be precise, we take the system's initial state $\rho^{AB}$ as the X state with the same local purity.
The time evolution of $\rho^{AB}$ is given by \eqref{sys state}.
The environment state $\rho^{A'B'}(t)$ can be attained by considering the total state $\rho^{ABA'B'}(t)$ and then tracing out $A'$ and $B'$.
We see that at any time both $\rho^{AB}(t)$ and $\rho^{A'B'}(t)$ are the sort of the X state with the identical local purity, which allows for an analytical computation of the geometric quantum correlation $G$.

Let us consider the following two examples.

\begin{figure}[bpth]
\begin{center}
  \includegraphics[width=0.4\textwidth]{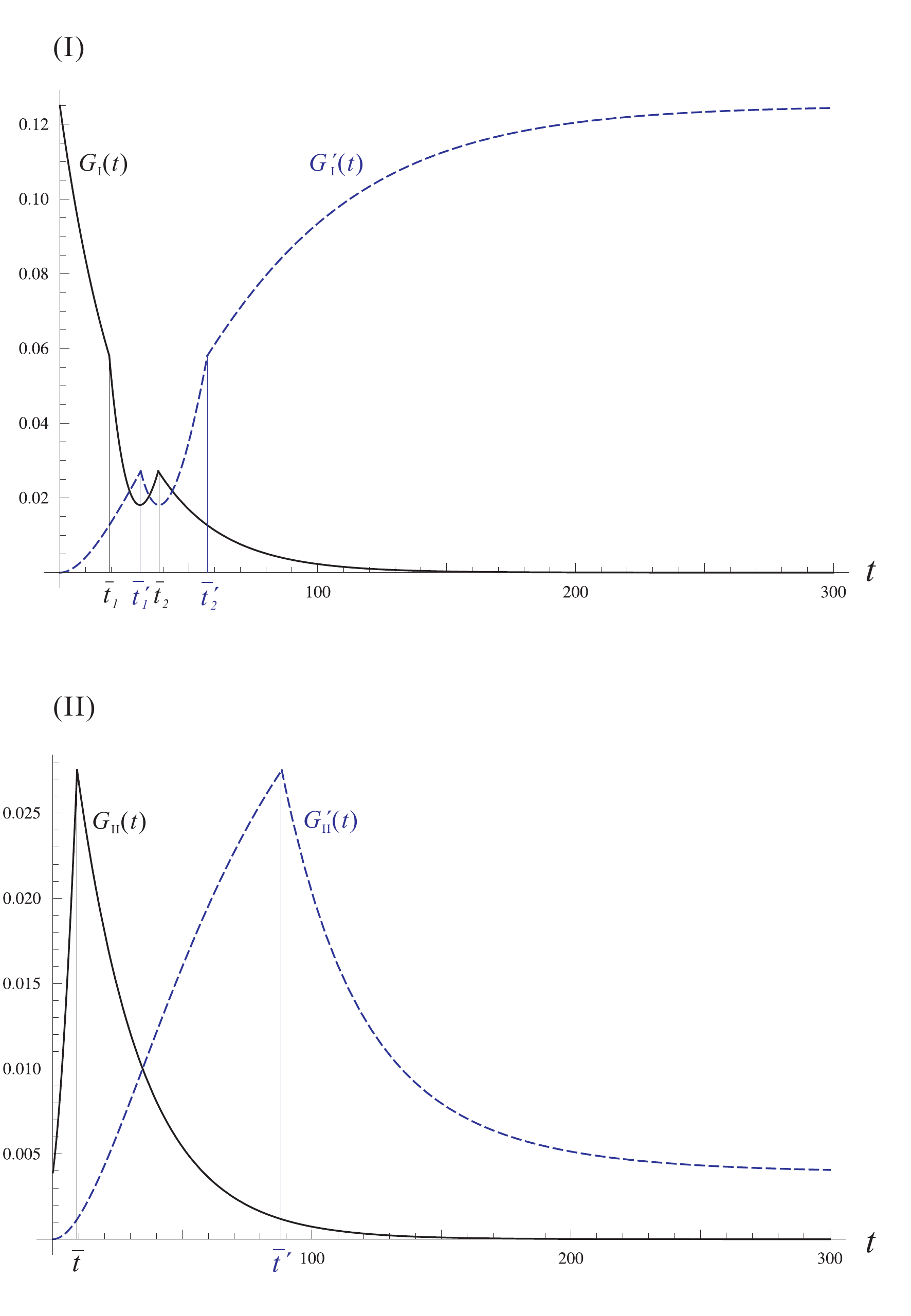} \\
  \caption{Plots of the quantum correlation evolutions $G(t)$ (solid line) and $G'(t)$ (dashed line) for states $\rho^{AB}_{\;\mathrm{I}}$ (upper panel) and $\rho^{AB}_{\;\mathrm{I\!I}}$ (lower panel) respectively. See text for details.} \label{fig:typical QD}
\end{center}
\end{figure}

Suppose that two 2-qubit X states $\rho^{AB}_{\;\mathrm{I}}$ and $\rho^{AB}_{\;\mathrm{I\!I}}$ are given by, in terms of the elements in the $R$ matrix,
\begin{equation}
  \rho^{AB}_{\;\mathrm{I}}:\quad
  \begin{cases}
    x_3=y_3=0.7949, \\
    T=\diag(0.4705,\,-0.5277,\,0.8947),
  \end{cases}
\end{equation}
and
\begin{equation}
  \rho^{AB}_{\;\mathrm{I\!I}}:\quad
  \begin{cases}
    x_3=y_3=0.6479,\\
    T=\diag(0.3926,\,-0.0772,\,0.0360),
  \end{cases}
\end{equation}
respectively.
The coupling strength $\kappa$ of the AD channel is set to be $0.02$.
We calculate $G(\rho_k^{AB}(t))$ and $G(\rho_k^{A'B'}(t))$ (or simply $G_k(t)$ and $G'_k(t)$) with $k=\mathrm{I},\mathrm{I\!I}$, and plot the results in Fig. \ref{fig:typical QD}. 
We observe several nontrivial aspects.

(i) The time evolution of quantum correlation, whether for system qubits $AB$ or for the environments $A'B'$, is neither monotone nor smooth.
There exist critical times on which the evolution rate exhibits a sudden change in behavior. The sudden change may even happen twice.
More importantly, the quantum correlation in the system state can increase during the time evolution, even if the state $\rho^{AB}$ is suffering a dissipative process.

(ii) By comparing $G_{\mathrm{I}}(t)$ with $G'_{\mathrm{I}}(t)$, we see that there is a symmetry between them. Roughly speaking, the evolution of $G_{\mathrm{I}}$ from $t=0$ to $t\to\infty$ is similar to the ``inverse'' evolution of $G'_{\mathrm{I}}$, namely from $t\to\infty$ to $t=0$. Precisely, we have the following relationship: $G_{\mathrm{I}}(0)=G'_{\mathrm{I}}(+\infty)$,
$G_{\mathrm{I}}(\bar{t}_1)=G'_{\mathrm{I}}(\bar{t}^{\,\prime}_2)$,
$G_{\mathrm{I}}(\bar{t}_2)=G'_{\mathrm{I}}(\bar{t}^{\,\prime}_1)$, and
$G_{\mathrm{I}}(+\infty)=G'_{\mathrm{I}}(0)$, where $\bar{t}_i$ and $\bar{t}^{\,\prime}_i$ ($i=1,2$) are the critical time for $G_{\mathrm{I}}(t)$ and $G'_{\mathrm{I}}(t)$, respectively (see Fig.\ref{fig:typical QD}).
In other words, there is a correspondence at the initial, critical and asymptotical point.
A similar situation occurs in the comparison of $G_{\mathrm{I\!I}}(t)$ and $G'_{\mathrm{I\!I}}(t)$.

(iii) The fact that $G(0)=G'(+\infty)$ means that the initial quantum correlation between the system qubits is transferred completely to the asymptotical quantum correlation between the ancillary qubits.

Let us make some remarks about the above observations and related topics.
 
For a 2-qubit system interacting with two independent environments, the dynamics and the sudden change behavior of quantum discord have been analyzed in Ref. \cite{Maziero.PhysRevA.80.044102.2009,
             Mazzola.PhysRevLett.104.200401.2010,
             Maziero.PhysRevA.81.022116.2010}.
Subsequently, these issues are discussed in \cite{Lu.QuanInfCompt.10.994.2010} in terms of another quantifier of quantum correlation, one-sided HS distance measure \cite{Dakic.PhysRevLett.105.190502.2010}. 
The results we present here are referring to the two-sided (i.e., symmetric) HS distance measure.
One significant phenomenon is that quantum correlation $G$ can increase under the influence of the separate independent dissipative environments.

Recently, some authors show that the quantum discord can increase under a local amplitude
damping channel with a CC input state \cite{Ciccarello.arXiv.1105.5551}.
A more general conclusion is drawn in \cite{Streltsov.arXiv.1106.2028}, that is,
any local channel which is non-unital and not semi-classical can in principle create quantum correlations, independently of the considered measure, out of a CC state.
Our work provides a concrete evolution process, in particular, with respect to the state $\rho_{\mathrm{I\!I}}^{AB}$, in which we see a rapid increase of the quantum correlation in system qubits. 
It should be noted that the input state $\rho_{\mathrm{I\!I}}^{AB}$ is a separable but not a CC state. Meanwhile, we also see that the quantum correlation of the environments undergoes a similar evolution: $G'_{\mathrm{I\!I}}(t)$ increases in the beginning and then changes suddenly to decrease.

It should be mentioned that quantum discord can increase in the non-Markovian environments \cite{Wang.PhysRevA.81.014101.2010,Fanchini.PhysRevA.81.052107.2010}, whereas the AD channel corresponds to a Markovian process.

Concerning the asymptotical behavior of $G(t)$, we see that 
$G(t)\propto\gamma^2(t)=e^{-\kappa t}$ after the second sudden change (if existing).
This behavior is qualitatively identical to the decay of the skew diagonal elements of the density matrix $\rho^{AB}(t)$, e.g., $\big(\rho^{AB}(t)\big)_{14}\propto e^{-\kappa t}$. 
It means that the asymptotical evolution of quantum correlation is closely related the coherence decay (in the case discussed above, they are qualitatively identical). 
Our discussion, from the viewpoint of symmetric distance quantifier,  provides an evidence for the robustness of quantum discord \cite{Werlang.PhysRevA.80.024103.2009} and the claim that ``almost all quantum states have nonclassical correlations'' \cite{Ferraro.PhysRevA.81.052318.2010}.

\section{Conclusion}

In conclusion, we have introduced in this paper a symmetric geometric measure of quantum correlation. 
By performing two-sided von Neumann measurements on bipartite state, we obtain a MICC state. The geometric measure is defined as the HS distance between the given state and the closest MICC state. 
For 2-qubit system, we simplify the optimization procedure considerably, that is, the two-sided optimization is reduced to the one-sided one.
Hence the numerical computation can be performed efficiently.
Moreover, the analytical results are available for some special class of 2-qubit states.

Using this quantifier, we have studied the dynamics of quantum correlation under the action of AD channel. 
We present the nontrivial aspects which may be exhibited during the time evolution: 
(i) the quantum correlation can increase; 
(ii) the quantum correlation can change suddenly, even twice. 
As for the environments, we see that the quantum correlation therein increases from zero at the beginning, and then evolves asymptotically to the value of the initial quantum correlation in the system qubits, after one or two sudden changes.

The geometric measure developed in this paper provides a symmetric viewpoint to study the quantum correlation.
It is also a measurement-oriented measure, since the minimal HS distance is referring to the MICC, rather general CC, states. Then a question follows: Is there more rigorous quantifier, which is referring to the general CC states? This question has been solved in the case of one-sided geometric measure \cite{Luo.PhysRevA.82.034302.2010}. However, it remains open for the two-sided measure.

\begin{acknowledgments}
This work was supported by National Nature Science Foundation of China, the CAS, and the National
Fundamental Research Program 2007CB925200.
\end{acknowledgments}

\appendix
\section{$G$ for X state with the same local purity}\label{appendix}

Let $|x_3|=|y_3|=r$ with $0<r<1$.
We assume that all $t_i$ is nonzero for the sake of simplicity.
Referring to \eqref{lambdaM}, we rewrite $\lambda_M(\vec{\ell}\,)$ as
\begin{equation} \label{lambdaM X state case3}
  \lambda_M(\vec{\ell}\,)=\frac{1}{2}\Big[r^2+\ell^{\prime\,2}
    +\sqrt{(r^2-\ell^{\prime\,2})^2+4r^2t_3^2\ell_3^2}\,\Big]+r^2\ell_3^2.
\end{equation}
Define the function $F(\vec{\ell}\,)$ as
\begin{equation}
  F(\vec{\ell}\,)=(x^2-\ell^{\prime\,2})^2+4r^2t_3^2\ell_3^2.
\end{equation}
In the following, we discuss two cases separately: $t_1=t_2$ and $t_1\neq t_2$.

\subsection{Case of $t_1=t_2$} \label{Sec:case t1 equal to t2}

When $t_1=t_2$, we have $\ell^{\prime\,2}=t_1^2-(t_1^2-t_3^2)\ell_3^2$.
So both functions $F(\vec{\ell}\,)$ and $\lambda_M(\vec{\ell}\,)$ depend only on $\ell_3^2$.
By taking derivative of $\lambda_M(\ell_3^2)$ with respect to $\ell_3^2$, and solving the equation $\frac{d\lambda_M(\ell_3^2)}{d(\ell_3^2)}=0$ for $\ell_3^2$, we have the following results.

If $t_1t_3>0$ and $r^2\in[t_3(t_1-t_3),\,t_1(t_1-t_3)]$, we have
\begin{equation} \label{solution:l3sq1 case1}
  \ell_3^2=\frac{-r^2+t_1(t_1-t_3)}{(t_1-t_3)^2}.
\end{equation}

If $t_1t_3<0$ and $r^2\in[-t_3(t_1+t_3),\,t_1(t_1+t_3)]$, we have
\begin{equation} \label{solution:l3sq2 case2}
  \ell_3^2=\frac{-r^2+t_1(t_1+t_3)}{(t_1+t_3)^2}.
\end{equation}
Some remarks are needed here. 

(i) The conditions for \eqref{solution:l3sq1 case1} and \eqref{solution:l3sq2 case2} come from two considerations: 
one is the non-negativity of $F(\ell_3^2)$; 
the other is the requirement of $\ell_3^2\in[0,1]$. 

(ii) In solving the equation $\frac{d\lambda_M(\ell_3^2)}{d(\ell_3^2)}=0$, it is assumed that $F(\ell_3^2)\neq0$. 
In fact, if $F(\ell_3^2)=0$, we have $\ell_3=0$ and $r^2=\ell^{\prime\,2}$. 
It follows that $\lambda_M=r^2$. 
We will see below that $r^2$ cannot be the maximal value of $\lambda_M(\ell_3^2)$.

(iii) Eqs. \eqref{solution:l3sq1 case1} and \eqref{solution:l3sq2 case2} require that $t_1\neq t_3$ and $t_1\neq -t_3$, respectively. 
In fact, if $t_1=t_3$, we have $t_1=t_2=t_3=t$ and $\ell^{\prime\,2}=t^2$. 
It follows from \eqref{lambdaM X state case3} that
\begin{equation*}
  \lambda_M(\vec{\ell}\,)=\frac{1}{2}\Big[r^2+t^2
    +\sqrt{(r^2-t^2)^2+4r^2t^2\ell_3^2}\,\Big]+r^2\ell_3^2.
\end{equation*}
It is a monotone increasing function for $\ell_3^2\in[0,1]$. 
Then $\lambda_M^{\max}=2r^2+t^2$ when $\ell_3^2=1$. 
This result can be contained in \eqref{lambdaM4}. Similarly for the case of $t_1=-t_3$.
This analysis also holds for the derivation process in the next Case.

Inserting \eqref{solution:l3sq1 case1} and \eqref{solution:l3sq2 case2} into \eqref{lambdaM X state case3} respectively, we get two candidates for $\lambda_M^{\max}$:
\begin{align}
   & \lambda_M^{(1)}=\frac{r^2[2t_1(t_1-t_3)-r^2]}{(t_1-t_3)^2}, \label{lambdaM1}\\
   & \lambda_M^{(2)}=\frac{r^2[2t_1(t_1+t_3)-r^2]}{(t_1+t_3)^2}. \label{lambdaM2}
\end{align}
We have to take the end points of $\ell_3^2$ into consideration, i.e., $\ell_3^2=0$ and $\ell_3^2=1$.
When $\ell_3^2=0$, we have $\lambda_M=\big[r^2+t_1^2+|r^2-t_1^2|\big]/2$. Let's define
\begin{equation} \label{lambdaM3}
  \lambda_M^{(3)}=\max\{r^2,\,t_1^2\}.
\end{equation}
When $\ell_3^2=1$, we have
\begin{equation} \label{lambdaM4}
  \lambda_M^{(4)}=2r^2+t_3^2.
\end{equation}
Combining the above analysis, we conclude as follows.

If $t_1t_3>0$ and $r^2\in[t_3(t_1-t_3),\,t_1(t_1-t_3)]$, the maximal $\lambda_M(\vec{\ell}\,)$ is given by
\begin{equation}
  \lambda_M^{\max}=\max\{\lambda_M^{(1)},2r^2+t_3^2,t_1^2\}.
\end{equation}

If $t_1t_3<0$ and $r^2\in[-t_3(t_1+t_3),\,t_1(t_1+t_3)]$, the maximal $\lambda_M(\vec{\ell}\,)$ is given by
\begin{equation}
  \lambda_M^{\max}=\max\{\lambda_M^{(2)},2r^2+t_3^2,t_1^2\}.
\end{equation}

\subsection{Case of $t_1\neq t_2$}

Let's prove the following lemma.
\begin{lemma} \label{lemma 2}
If $t_1\neq t_2$, then at least one of $\ell_i$ is zero.
\end{lemma}

With $\lambda_M(\vec{\ell}\,)$ given by \eqref{lambdaM X state case3}, we introduce Lagrange multiplier $\mu$ and take partial derivative of $\lambda_M(\vec{\ell}\,)+\mu(\ell^2-1)$ with respect to $\ell_i$. Three equations follow.
\begin{align}
    & \ell_1\bigg[t_1^2-\frac{t_1^2(r^2-\ell^{\prime\,2})}{\sqrt{F}}+2\mu\bigg]=0,
      \label{eq:partial l1 case3} \\
    & \ell_2\bigg[t_2^2-\frac{t_2^2(r^2-\ell^{\prime\,2})}{\sqrt{F}}+2\mu\bigg]=0,
      \label{eq:partial l2 case3} \\
    & \ell_3\bigg[t_3^2+\frac{t_3^2(r^2+\ell^{\prime\,2})}{\sqrt{F}}+2r^2+2\mu\bigg]=0.
      \label{eq:partial l3 case3}
\end{align}
Assume that all $\ell_i$ are nonzero. If so, we can delete $\ell_i$ in each equation. By noting that all $t_i$ are nonzero (as assumed at the beginning of this Subsection), we see from \eqref{eq:partial l1 case3} and \eqref{eq:partial l2 case3} that the Lagrange multiplier $\mu$ must be zero. Then Eqs. \eqref{eq:partial l1 case3} or \eqref{eq:partial l2 case3} reduce to $1-(r^2-\ell^{\prime\,2})/\sqrt{F}=0$. It follows that $\ell_3=0$, which contradicts the assumption. Lemma \ref{lemma 2} is proved.

Subsequently, we will discuss one by one the measurements allowed by Lemma \ref{lemma 2}. In each case, we obtain a candidate for $\lambda_M^{\max}$. The largest one is what we want.

\paragraph{If $\ell_1=0$ and $\ell_2,\ell_3\neq0$.} It follows that
$\ell^{\prime\,2}=t_2^2-(t_2^2-t_3^2)\ell_3^2$, and that both $\lambda_M(\vec\ell\,)$ and $F(\vec\ell\,)$ are the functions of $\ell_3^2$ only. By the approach similar to that presented in Section \ref{Sec:case t1 equal to t2}, we have the following results.

If $t_2t_3>0$ and $r^2\in[t_3(t_2-t_3),\,t_1(t_2-t_3)]$, we have
\begin{equation} \label{lambdaM5}
  \lambda_M^{(5)}=\frac{r^2[2t_2(t_2-t_3)-r^2]}{(t_2-t_3)^2}.
\end{equation}

If $t_2t_3<0$ and $r^2\in[-t_3(t_2+t_3),\,t_1(t_2+t_3)]$, we have
\begin{equation} \label{lambdaM6}
  \lambda_M^{(6)}=\frac{r^2[2t_2(t_2+t_3)-r^2]}{(t_2+t_3)^2}.
\end{equation}

\paragraph{If $\ell_2=0$ and $\ell_1,\ell_3\neq0$.} In this case, $\ell^{\prime\,2}=t_1^2-(t_1^2-t_3^2)\ell_3^2$. The results are very similar to that presented in the last paragraph, that is:

If $t_1t_3>0$, we have
\begin{equation} \label{lambdaM7}
  \lambda_M^{(7)}=\frac{r^2[2t_1(t_1-t_3)-r^2]}{(t_1-t_3)^2}.
\end{equation}

If $t_1t_3<0$, we have
\begin{equation} \label{lambdaM8}
  \lambda_M^{(8)}=\frac{r^2[2t_1(t_1+t_3)-r^2]}{(t_1+t_3)^2}.
\end{equation}

Note that $\lambda_M^{(7)}$ and $\lambda_M^{(8)}$ have the same form as $\lambda_M^{(1)}$ and $\lambda_M^{(2)}$ (given in \eqref{lambdaM1} and \eqref{lambdaM2}) respectively.

\paragraph{If $\ell_3=0$.} Here it is not required that both $\ell_2$ and $\ell_3$ are nonzero. Inserting $\ell_3=0$ into the expression of $\lambda_M(\vec{\ell}\,)$, i.e., Eq. \eqref{lambdaM X state case3}, we have
\begin{equation*}
    \lambda_M(\vec{\ell}\,)
      =\frac{1}{2}\big[r^2+(t_1^2\ell_1^2+t_2^2\ell_2^2)
        -|r^2-(t_1^2\ell_1^2+t_2^2\ell_2^2)|\big].
\end{equation*}
The maximal value of the above expression is given by
\begin{equation} \label{lambdaM9}
  \lambda_{M}^{(9)}=\max\{r^2,t_1^2,t_2^2\}.
\end{equation}

\paragraph{If $\ell_3=1$.} In this case, $\ell_1=\ell_2=0$. It easily follows that
\begin{equation} \label{lambdaM10}
  \lambda_M^{(10)}=2r^2+t_3^2.
\end{equation}

It is not difficult to see that the above four cases cover all allowed measurements. We summary the results obtained in this Subsection in Table \ref{table}.

\begin{center}
\begin{table}
\caption{Summary of the results in Section \ref{Sec:X states x3 eq y3}}\label{table}
\begin{ruledtabular}
\begin{tabular}{lll}
              $\vec{\ell}$   &   Conditions   &   $\lambda_M$ \\
                  \hline\vspace{-1em}\\ \vspace{5pt}
    $\ell_1=0$, and   &  $t_2t_3>0$, $r^2\in\mathrm{Interval}_1$
                      \footnote{$\mathrm{Interval_1}=[t_3(t_2-t_3),t_2(t_2-t_3)]$, $\mathrm{Interval_2}=[-t_3(t_2+t_3),t_2(t_2+t_3)]$, $\mathrm{Interval_3}=[t_3(t_1-t_3),t_1(t_1-t_3)]$, $\mathrm{Interval_4}=[-t_3(t_1+t_3),t_1(t_1+t_3)]$.}
                   &  $\lambda_M^{(5)}$,  \eqref{lambdaM5} \\
  $\ell_2,\ell_3\neq0$   &  $t_2t_3<0$, $r^2\in\mathrm{Interval}_2$
                   &  $\lambda_M^{(6)}$,  \eqref{lambdaM6} \vspace{5pt}\\
                 \hline\vspace{-1em}\\ \vspace{5pt}
    $\ell_2=0$, and   &  $t_1t_3>0$, $r^2\in\mathrm{Interval}_3$
                   &  $\lambda_M^{(7)}$,  \eqref{lambdaM7} \\
  $\ell_1,\ell_3\neq0$   &  $t_1t_3<0$, $r^2\in\mathrm{Interval}_4$ &  $\lambda_M^{(8)}$,  \eqref{lambdaM8} \vspace{5pt}\\
                 \hline\vspace{-1em}\\ \vspace{5pt}
  $\ell_3=0$          &
                   & $\lambda_M^{(9)}$,  \eqref{lambdaM9} \\
                 \hline\vspace{-1em}\\ \vspace{5pt}
  $\ell_3=1$          &
                   & $\lambda_M^{(10)}$,  \eqref{lambdaM10}  \\
\end{tabular}
\end{ruledtabular}
\end{table}
\end{center}

Now let's use an example to show how to calculate $\lambda_M^{\max}$. Given an X state $\rho^{AB}$ with $|x_3|=|y_3|=r$, we write its $R$ matrix. If we see that $t_2t_3>0$ and $t_1t_3>0$, then we face the following possibilities:
\begin{itemize}
\item[(i)] If $r^2\in\mathrm{Interval}_1$ and $r^2\in\mathrm{Interval}_3$, then $\lambda_M^{\max}=\max\{\lambda_M^{(5)},\lambda_M^{(7)},\lambda_M^{(9)},\lambda_M^{(10)}\}$.
\item[(ii)] If $r^2\in\mathrm{Interval}_1$ and $r^2\not\in\mathrm{Interval}_3$, then $\lambda_M^{\max}=\max\{\lambda_M^{(5)},\lambda_M^{(9)},\lambda_M^{(10)}\}$.
\item[(iii)] If $r^2\not\in\mathrm{Interval}_1$ and $r^2\in\mathrm{Interval}_3$, then $\lambda_M^{\max}=\max\{\lambda_M^{(7)},\lambda_M^{(9)},\lambda_M^{(10)}\}$.
\item[(iv)] If $r^2\not\in\mathrm{Interval}_1$ and $r^2\not\in\mathrm{Interval}_3$, then $\lambda_M^{\max}=\max\{\lambda_M^{(9)},\lambda_M^{(10)}\}$.
\end{itemize}
Then quantum discord $G(\rho)$ is obtained by inserting $\lambda_M^{\max}$ into \eqref{geo measure}.


\bibliography{GeoMeasureQD}

\clearpage
\end{document}